\documentclass[12pt]{article}
\usepackage[T1]{fontenc}
\usepackage{psfig}
\usepackage{bookman}
\usepackage{latexsym}
\usepackage{amssymb}
\usepackage{amsbsy}
\usepackage{cite}
\begin{document} 
  
  \title{Evolution of the Density Contrast with Particle Production in a
 Gauge-Invariant Formalism}
 \author{ M. de Campos $^{(1)}$,  N.A.Tomimura$^{(1)}$}
\maketitle
   
   \footnote{
{ \small \it $^{(1)}$ Instituto de F\'\i sica \vspace{-0.2cm}}\\
{ \small \it Universidade Federal Fluminense  \vspace{-0.2cm}}\\
{\small \it Av. Gal. Milton Tavares de Souza s/n.$\!\!^\circ$, 
                  \vspace{-0.2cm} }}    
      \begin{abstract}
In this work we derive the evolution equation for the density contrast
considering open system cosmology, where the influence of adiabatic particle
production process on the dynamic of a homogeneous and isotropic is
investigated within a manifestly covariant formulation. As application we
derive the solution for two sources, one of them is a generalization of
Prigogine's model.  Then we establish the condition for the reach of the
non-linear regime for the density contrast which turns out to be a necessary
condition for the structure formation.
\end{abstract}
%
%
%
%
%
%SECAO 1
%
%
%
%
%
\section{Introduction}
The mathematical theory of perturbations in homogeneous, isotropic models has been worked over many times in the 
literature \cite{0}-\cite{Kodama}.  Nevertheless, troubling questions still remain about the physical interpretation of density perturbations 
at early times when the perturbation is larger than the particle horizon, which will here mean when the time for light
to travel a characteristic wavelength of the perturbation is larger than the instantaneous expansion time.  These questions 
are particularly relevant to attempt to explain the origin of perturbations which eventually give rise to galaxy 
formation.

The study of perturbations of an expanding universe would appear to be
hopelessly gauge dependent.  However,  Bardeen \cite{1} gave an important contribution to define
perturbations that are  gauge invariant non-geometrical
quantities defined with respect to a particular chart.  Thus a gauge-independent theory of 
cosmological perturbations is possible. However, Bardeen's paper is difficult to understand and to apply.

Ellis and Bruni \cite{2} gave an alternative approach to gauge invariant cosmological perturbations with basic variables
more closely related to the physical quantities.  They look for variables
which vanish in the background. 
 Quantities  of this kind are always gauge independents.  The gauge invariant key variable is the comoving
 fractional spatial gradient of the energy density.
 
 The purpose of this paper is to discuss an approach to cosmological perturbations which are thermodynamically oriented 
 gauge-invariant description of scalar cosmological perturbations around a
 flat homogeneous universe with particle production.  All basic
 variables have their physical meaning on ``comoving hyper-surfaces'' , orthogonal to the matter world lines.  
 Thermodynamic perturbations relations in Minkowski space-time turn-out to be valid in the expanding universe, provided
 all perturbed quantities are replaced by their gauge-invariant counterparts.
 
 The characteristic feature of this model is the use of an explicit phenomenological balance law for the particle number
 with a non-zero source term together with the equation for the cosmic scale
 factor in a perfect fluid cosmology that includes a pressure creation that
 behaves like a  viscous pressure, 
 responsible for the particle production.
 %
 %
 %SECAO 2
 %
 %
 \section{Open system thermodynamics}
 In the presence of matter creation, the appropriate analysis is performed in
 the context of thermodynamics of  open systems \cite{14}. This consideration 
 leads to an extension of thermodynamics as associated with cosmology to reinterpretation in Einstein's equations of 
 the matter stress  energy tensor. In the case of adiabatic transformation, the thermodynamical energy conservation leads
 to 
                 \begin{equation}
        d(\rho\,V)+P_{{{th}}}{dV}-{\frac {Ed(nV)}{n}} = 0  \, ,    
                        \end{equation}
  where $ n = \frac{N}{V} $ is the particle number density and $ E = \rho + P_{th} $ is the enthalpy per unit volume. $ N $ is the particle number
  and $ P_{th} $ is the thermodynamical pressure.
  In such a transformation, the ``heat'' received by the system is due entirely to the change of the number 
  of particles.  In our cosmological context, this change is due to the transfer of energy from gravitational field.
  Hence, the creation of matter acts as a source of internal energy.
  
  The cosmic medium will be described by the energy momentum tensor
             \begin{equation}
               {T}^{\alpha\,\beta}=\epsilon\,{u}^{\alpha}{u}^{\beta}+P{h}^{\alpha\, \beta} \, ,
                   \end{equation}
  where $\epsilon $ is the energy-density, $ h^{\alpha \beta }$ is the
 projection tensor and
 $ u^{\alpha}$  is the four-velocity. $ P $ is the total pressure that 
  includes the thermodynamical pressure and the creation pressure \cite{13}, namely,
  \begin{equation}
       P = P_{th} + \tilde{P} \, .
  \end{equation} 
  The creation process will be considered adiabatic, which here means a constant entropy per particle.  Under this
 condition the equilibrium entropy per particle does not change as it does in
 dissipative process.  Instead, one can associate a ``viscous pressure'' to the
 particle production source, namely \cite{13}
                 \begin{equation}
                      \tilde{P} =-{\frac {\left ({\it {\rho} }+P_{{{\it {th}}}}\right )\Psi}{n\theta}} \, ,
                        \end{equation}
  where $ \Psi $ is the rate of particle production.
  
  Einstein's field equations imply the equations of motion
  \begin{equation}
  h^{\alpha } _{\beta } T^{\gamma \beta } _{; \gamma} = 0 \, ,
  \end{equation}
  and the energy balance
  \begin{equation}
  u_{\beta} T^{\alpha \beta }  _{; \alpha} = 0 \, .
  \end{equation}
  
  Additionally, according to the second law of thermodynamics, the only particle number variations admitted are such that
                 \begin{equation}
                   \dot{n} + n \theta = \Psi \, .
                    \end{equation} 
  
  The general problem is split into a zeroth-order background universe and first order perturbation about this background.
    The zeroth-order is assumed to be homogeneous, isotropic and spatially flat. The corresponding line element is 
             \begin{equation}
                ds^{2} =-dt^2 +  R(t)^{2} \gamma_{a b} dx ^{a} dx ^{b} \, ,
               \end{equation} 
  with the cosmic scale factor  $ R(t) $ as the only relevant metric quantity. Consequently, Einstein's field equations
  reduce to
               \begin{eqnarray}
                  3 ( \frac{\dot{R}}{R})^{2} = \epsilon \\
                      2 (\frac{\dot{R}}{R})^{\cdot } = -(\epsilon + P) \, .
                        \end{eqnarray}
  The dot denotes the derivative with respect to time. The surfaces $ t = const. $ are the space like surfaces
   of constant
  curvature and $ \gamma _{a b} $ is the 3-space metric.
  Consequently, the energy balance equation (6) becomes
  \begin{equation}
  \dot{\epsilon } = -3 \frac{\dot{R}}{R} (\epsilon +P ) \, .
  \end{equation} 
  Considering the particle number density and the temperature as are our basic variables we can infer the state equations
  \begin{eqnarray}
  P_{th} = P_{th} (n,T) \\
  \epsilon = \epsilon (n,T) \\
  \sigma = \sigma (n,T) \, ,
  \end{eqnarray}
 where $ \sigma $ is the entropy per particle.
 Making use of Gibbs relation and equation (14), it is straightforward to find the general relation  
 \begin{equation}
 \frac{\partial \epsilon}{\partial n} = \frac{\epsilon + P_{th}}{n} - \frac{T}{n} \frac{\partial P_{th}}{\partial T} \, .
 \end{equation} 
 Equation for evolution of the temperature is calculated considering relations (11) and (15), namely
 \begin{equation}
 \dot{T} = (\frac{\partial \epsilon}{\partial T})^{-1} \{ 3\frac{\dot{R}}{R} (\tilde{P} + \frac{\partial P_{th}}
 {\partial T} T) +\Psi (\frac{\epsilon + P_{th}}{n} - \frac{T}{n} \frac{\partial P_{th}}{\partial T})\} \, .  
 \end{equation}
We can define the expression for the  sound velocity as:
\begin{equation}
v_{s}^2 = \frac{\partial P_{th}}{\partial \epsilon} \nonumber
\end{equation}
that can be reformulated using equation (15):
\begin{equation}
v_{s}^2 = \frac{n}{\epsilon + P_{th}}\frac{\partial P_{th}}{\partial n} +
\frac{T{[\frac{\partial P_{th}}{\partial
      T}]}^2}{(\epsilon+P_{th})\frac{\partial \epsilon }{\partial T}} \, .
\end{equation}
 Taking into account the relation
 \begin{equation}
 \nonumber \dot{P_{th}} = \frac{\partial P_{th}}{\partial n} \dot{n} + \frac{\partial P_{th}}{\partial T} \dot{T}
 \end{equation}
 and equations (7), (16) and (17), evolution of the pressure reads
 \begin{eqnarray}
 &\dot{P_{th}}& = -3\frac{\dot{R}}{R} (\epsilon + P_{th}) \{v_{s}^2 +\frac{\frac{\partial P_{th}}
 {\partial T}}{\frac{\partial \epsilon}{\partial T}} \frac{\tilde{P}}{\epsilon + P_{th}} \}  
 \frac{\Psi}{n} (\epsilon + P_{th}) \\
&\{& v_{s}^2- \frac{\frac{\partial P_{th}}{\partial T}}
 {\frac{\partial \epsilon}{\partial T}}\} \, .
 \end{eqnarray}
 Finally the time evolution of the entropy density can be found by using Gibbs relation and equations (7) and (16), namely
 \begin{equation}
 nT\dot{\sigma} = \frac{\epsilon + P_{th}}{n} \{  3n\frac{\dot{R}}{R} z - \Psi\}
 \end{equation}
 where
 \begin{equation}
 z = -\frac{\tilde{P}}{\epsilon + P_{th}} \, .
 \end{equation}                        
 This relation provides a direct connection between the particle production rate and the expansion of the Universe.
  Considering the
 process as adiabatic  together with (7) yields
 \begin{equation}
 z = \frac{\frac{\dot{N}}{N}}{\theta } \, .
 \end{equation}
 We sum up this section by writing the time evolution of the particle density, temperature, energy density and pressure
 in the adiabatic case, it is done by using (7), (16) and (19)
 \begin{eqnarray}
 \dot{n} = -3\frac{\dot{R}}{R} (1-z) \, , \\
 \dot{T} = =3\frac{\dot{R}}{R}T (\frac{\frac{\partial P_{th}}{\partial
 T}}{\frac{\partial \epsilon}{\partial T}}) \, , \\
 \dot{\epsilon} = -3 \frac{\dot{R}}{R} (\epsilon + P) \, , \\
 \dot{P_{th}} = -3\frac{\dot{R}}{R} (\epsilon + P) v_{s}^2 \, .
 \end{eqnarray}
 %OBS: Bulk viscosity and matter creation generate cosmological models with sam%e physical properties , but the identity
% between the two process is not generically valid.
% Cosmic dynamics generated by each can be mimicked by the other one.  However %the thermodynamic features are quite 
% different \cite{3}.
 %
 %
 %
 %
 %SECAO 3
 %
 %
 %
 %
 %
 \section{The perturbative equations}
 We write the metric as
 \begin{equation}
 ds^2 = (g_{\alpha \beta} + \delta g_{\alpha \beta}) dx^{\alpha} dx^{\beta} \, ,
 \end{equation}
 where $g_{\alpha \beta}$ is the background metric tensor and 
 \begin{eqnarray}
\delta g_{ab} = A\delta_{ab} + B_{,ab}\, ,     \\
\delta g_{a4} = F_{,4}  \, ,  \\
\delta g_{44} = E \, .
 \end{eqnarray}
 The quantities $A, B, F $ e $E $ are space and time dependent scalars hence
 we are interested in scalar perturbations. 
 
 From normalization condition, $u_{\mu }u^{\mu } = 1$ and the line element (29) the four velocity in first order is given by
 \begin{eqnarray}
 \hat{u_{4}} = \hat{u^{4}} = \frac{R^2 E}{2} \, ,\\ 
 R^2\hat{u^{a}} + R^2 F_{,a} = \hat{u_{a}} \equiv v_{,a} \, ,
 \end{eqnarray}
where the hat indicates a  perturbative
 quantities.
 Perturbing to first order the equation which gives the conservation of particle density $(nu^{\alpha})_{;\alpha} = 
 \Psi$, the result is
 \begin{equation}
 (\frac{\hat{n}}{n})^{\cdot } + R^{-2} \nabla v -\nabla F + \frac{\dot{f}}{2} = \frac{\hat{\Psi }}{n}
 - \frac{\Psi}{n} (\frac{\hat{n}}{n} + \frac{R^2 E}{2}) \, .
\end{equation}
 
 The energy balance and momentum balance equations to first order are respectively:
 \begin{eqnarray}
 \dot{\hat{\epsilon}} +3\frac{\dot{R}}{R}(\hat{\epsilon} + \hat{P}) + (R^{-2}\nabla v + \frac{\dot{f}}{2} - \nabla F)
 (\epsilon + P) = 0 \, , \\
 \hat{P}_{,\mu} + \dot{P}v_{,\mu}+(\epsilon +
 P)\dot{v_{,\mu}}-R^{2}\frac{\epsilon + P}{2} E_{,\mu} =0 \, .
 \end{eqnarray}
 To complete the set of basic equations we need to linearize field
 equations, namely \cite{Kodama}:
 \begin{equation}
 \hat{G}^{4} _{4} = R^{-2}\nabla A
 -\frac{\dot{R}}{R}\dot{f}+2\frac{\dot{R}}{R}\nabla F -3\dot{R}^2 E =
 -\hat{\epsilon} \, , \\
\end{equation}
\begin{eqnarray}
& \hat{G}_{4\alpha}& = -\dot{A}_{,\alpha} -R\dot{R} E_{,\alpha } - (\dot{R}^2
+ 2R \ddot{R}) F_{,\alpha} \\
  &=& -(\epsilon +P)v_{,\alpha} + R^2 PF_{,\alpha} \, . \nonumber
 \end{eqnarray}
%
%
%
%
%Secao
%
%
%
%
\section{Density contrast evolution equation}
Variables representing perturbations introduced in the preceding section
modify their values under gauge transformations. In the
linear perturbation theory it is necessary only to consider infinitesimal
 transformation, namely
\begin{equation}
x^{' \alpha} = x^{\alpha }- \xi ^{\alpha} \, .
\end{equation}
Zimdahl shows in his  work \cite{4} how to describe cosmological perturbations thermodynamically oriented by suitable 
quantities which are
invariant under transformations (40).  These objects have an obvious physical meaning and
 obey reasonable equations.
The behavior of scalars under transformation (40) follow
\begin{equation}
S^{'} (x)-S(x)= \xi^{n} S_{,n}\, .
\end{equation}
Therefore the scalar quantities of the preceding section transform as
\begin{eqnarray}
A^{'} = A + 2\frac{\dot{R}}{R} \xi^{0} \\
B^{'} = B + 2R^{-2} \xi \\
F^{'} = F + R^{-2} \xi^{0}+(R^{-2} \xi)_{,0} \\
E^{'} = E - 2R^{-2} \xi_{0} ^{0} \\
v^{'} = v - \xi^{0} \\ 
\hat{n}^{'} = \hat{n} -3 \frac{\dot{R}}{R}n \xi^{0} (1-z) \\
\hat{T}^{'} = \hat{T} -3 \frac{\dot{R}}{R} n \xi ^{0} T (\frac{\frac{\partial
    P_{Th}}{\partial T}}{\frac{\partial \epsilon}{\partial T}}) (1-z) \, .
\end{eqnarray}
The basic variables that we shall take are 
\begin{equation}
s = \frac{\hat{n}}{n} -3 \frac{\dot{R}}{R} v (1-z)
\end{equation}
and 
\begin{equation}
r = \frac{\hat{T}}{T} -3 \frac{\dot{R}}{R} v \frac{\frac{\partial
    P_{th}}{\partial T}}{\frac{\partial \epsilon}{\partial T}} (1-z)\, .
\end{equation}
A gauge invariant gravitational field variable from the point of view of the
matter \cite{5} is:
\begin{equation}
q = \frac{3}{2} A + 3 \frac{\dot{R}}{R}v \, .
\end{equation}
It will turn out that the variables $q $ , $ r$ and $ s$ are sufficient for
the gauge-invariant description of the system of perturbation equations of the
preceding section.

The energy density perturbation and pressure perturbation are gauge
invariantly characterized by:
\begin{eqnarray}
\hat{\epsilon}_{c} = \hat{\epsilon} -3\frac{\dot{R}}{R}v (\epsilon +\tilde{P})\\
\hat{P}_{c} = \hat{P}_{th} -3 \frac{\dot{R}}{R} v (\epsilon + \tilde{P})
v_{s}^2 \, .
\end{eqnarray}
We can re-write the equation (49) by using (50):
\begin{equation}
\hat{\epsilon}_{c} = (\epsilon + P_{th})s +Ta \, ,
\end{equation}
where
\begin{equation}
a = \frac{\partial \epsilon}{\partial T} r - \frac{\partial P_{th} }{\partial
  T}s \, .
\end{equation}

Introducing the gauge invariant variables, the particle number balance (35) may
be written as
\begin{eqnarray}
&\dot{s}&+(1-z)\dot{q} +\frac{3}{2} \gamma \frac{\dot{R}}{R}
\frac{\hat{\epsilon}_{c}}{\epsilon + P_{th}} - \frac{R}{3\dot{R}} \frac{\kappa
  ^2}{R^2} q \\
&+& 9 \frac{\dot{R}^2}{R^2} z (1-z) v (1+\frac{\gamma}{2}) - 
3\frac{\dot{R}}{R} v \dot{z} + 3\frac{\dot{R}}{R} z s = \frac{\hat{\Psi}}{n}
\, . \nonumber
\end{eqnarray}
To find the last result it is necessary to use the perturbative field equations
(38), (39), the field equation
\begin{equation}
(\frac{\dot{R}}{R})^{\cdot } = -\frac{3}{2}H^2 \gamma (1-z)
\end{equation}
and assumes the spatial dependence $e^{\bf{ik\cdot r}}$ with constant comoving
wavector $k $ for all first-order quantities.

When equation (56) is used into the energy balance (36) and the  pressure
balance (37), they become respectively
\begin{equation}
\frac{(aR^3)^{\cdot }TR^{-3}}{\epsilon + P_{th}} + \frac{zk^2 q}{3R\dot{R}} +
\frac{\hat{\psi }}{n} = -3Hz\frac{\hat{P_{th}}-\frac{\gamma
    }{2}\epsilon}{\epsilon +P_{th}} - \frac{3H\hat{\tilde{P}}}{\epsilon + P_{th}}
  \end{equation}
and
\begin{eqnarray}
&sv_{s}^2& -v\dot{z}+3Hz(1-z)v(1+v_{s}^2) +\frac{\hat{\tilde{P}}}{\epsilon
  +P_{th}} +\frac{R}{3\dot{R}}\dot{q}(1-z) \\
&+&T\frac{\frac{\partial
    P_{th}}{\partial T}}{\frac{\partial \epsilon }{\partial T }}
\frac{a}{\epsilon +P_{th}}= 0 \, . \nonumber
\end{eqnarray}
Substituting equation (59) into (56) it becomes
\begin{eqnarray}
&\dot{s}&-3Hsv_{s}^2+9H^2 z(1-z)v(\frac{\gamma
  }{2}-v_{s}^2)-\frac{3H\hat{\tilde{P}}}{\epsilon +
  P_{th}} \\  \nonumber
&-&\frac{3HT}{\epsilon + P_{th}} a\frac{\frac{\partial
    P_{th}}{\partial T}}{\frac{\partial \epsilon }{\partial T }}
 + \frac{3}{2}\gamma H\frac{\hat{\epsilon_{c}}}{\epsilon + P_{th}}
  -\frac{k^2q}{3R\dot{R}}+3Hzs = \frac{\hat{\Psi }}{n} \, .
\end{eqnarray}
Equation (60) can be written in terms of  gauge invariant variables using
equation (55), namely:
\begin{eqnarray}
&\dot{s}&-3H\frac{1-z}{\epsilon + P_{th}}(\hat{P_{c}}-\frac{\gamma }{2}
\hat{\epsilon}_{c})-(1-z)\frac{k^2 q}{3R\dot{R}} \\ \nonumber
&+&3Hzs = -\frac{T}{\epsilon +
  P_{th}R^3 }(R^3 a)^{\cdot } \, .
\end{eqnarray}
We next focus our attention on the adiabatic case that is equivalent to consider $a = 0 $
\cite{4}, consequently equation (61) assumes the form
\begin{equation}
\dot{s} -3H(1-z)(v_{s}^2 -\frac{\gamma }{2})s-(1-z)\frac{k^2q}{3R\dot{R}}+3Hzs
= 0
\end{equation}
where the variable $s $ is given by
\begin{equation}
s = \frac{\hat{\epsilon}}{\epsilon + P_{th}} \, .
\end{equation}
A straightforward manipulation of equation (62) gives
\begin{eqnarray}
&\ddot{s}&+\{-\frac{3}{2}\{H(1-z)(2v_{s}^2 -\gamma )\}
+3Hz+(\frac{\ddot{R}}{\dot{R}} +\frac{\dot{R}}{R})+\frac{\dot{z}}{1-z} \}
  \dot{s} \\ \nonumber
&+&\{-\frac{3}{2}\{H(1-z)(2v_{s}^2-\gamma
  )\}^{\cdot}+3(Hz)^{\cdot}+\{3\frac{Hz}{1-z} \\ \nonumber
&-&\frac{3}{2}H(2v_{s}^2 -\gamma
  ) \} \{(1-z)(\frac{\ddot{R}}{\dot{R}}+\frac{\dot{R}}{R})+\dot{z}
  \}+\frac{k^2}{R^2}z\}s \\ \nonumber
&-&\frac{k^2}{R^2}v\dot{z} = 0 \, .
\end{eqnarray}
This equation governs  the behavior of the gauge invariant variable $s$ that
 is proportional to the density contrast, in an universe with particle production
at the expenses of the gravitational field.  

We recover the well known time evolution for
the  density contrast when the particle production is absent, namely
\begin{equation}
\ddot{s}+\frac{4}{3t}\dot{s} -\frac{2}{3t^2}s = 0 \, ,
\end{equation}
where the scale factor of the background, $R \propto t^{\frac{2}{3}}$ were used.
%
%
%
%Section
%
%
%
%
%
\section{Cosmological environment for particle production}
To integrate equation (64) the scale factor $R(t)$ must be specified.  This is
done when the source of the particle production $\Psi $ is chosen.
 
Next, we study the time evolution for the  perturbation of the matter density
 taking the source  of the particle production given by $\Psi = 3n\beta H $ and $ \Psi
 = \alpha H^{\frac{2}{\nu}}$.  Our purpose is to obtain the conditions for the
 reach of the non-linear regime in a  ``viscous universe'' with particle production. 
%
%
%SUBSECAO
%
%
\subsection{Source proportional to the product of the Hubble function and the
 particle number density}
Lima et al. \cite{6} gave a reasonable justification to pick up $\Psi $  as $\Psi =
3n\beta H $ , where $\beta  $ is a positive constant .
We can provide explicit solution for the time dependence  of the expansion
factor in the presence of particle creation by using the continuity equation
\begin{equation}
\dot{n} + 3Hn = \Psi \, ,
\end{equation}
 state equation
\begin{equation}
P_{th}= (\nu -1)\rho
\end{equation}
and the field equation
\begin{equation}
              R\ddot{R}+\Delta (k + \dot{R^{2} ) }=0 \, ,
               \end{equation}
where
     \begin{equation}
          \Delta = \frac{3}{2} \nu (1- \beta )-1 \, .
                     \end{equation}
We obtain, taking $\kappa = 0 $ and $\beta \neq 1$ :
         \begin{equation}
                  R = R_{0}(\frac{t}{t_{0}})^{\frac{1}{\Delta +1}} \, .
                       \end{equation}
The subscript $ 0 $ alludes to present time.  Note that for $\beta =1 $ the dominant
energy condition is violated \cite{12}.

Substituting the scale factor (70) and the respective source of particle
production in equation (64), we obtain
\begin{equation}
               \ddot{s}+C_{1} \dot{s}- C_{0} s = 0 \, ,
                    \end{equation}
where
         \begin{equation}
             C_{1} = \frac{2(3\beta +2)}{3\nu (1-\beta) t}
                      \end{equation}
and
                            \begin{equation}
C_{0} =-\frac{-2}{t^2} -\frac{4}{3\nu (\beta -1)^2 t^2}
\{3\beta^2-1+4\beta-\frac{4\beta }{\nu }\}  \, .
                          \end{equation} 
To solve equation (71) we neglect terms that contain the sound velocity and consider the limit of the large wavelengths.                   
Solutions  for  equation (71) are 

\begin{eqnarray}
s_{-}(t) =  C_{1} t^{-1+\frac{2\beta }{\nu (\beta -1)}} \\
s_{+}(t) = C_{2}t^{\frac{4}{3\nu (\beta -1)}+2} \, .
\end{eqnarray}
In the absence of particle production, $\beta = 0 $, relations (74) and (75) become the
usual density contrast for Friedmann models without creation \cite{7}.
Although the decreasing mode can be important in some circumstances, we shall
hereafter mainly deal only with the increasing mode. It is responsible for the
formation of cosmic structure in the gravitational stability picture.   Taking
into account the decreasing mode the universe would not have been homogeneous in
the past.  Besides, a growing mode that starts to grow just after the end of radiation era,
Peebles argued that the contribution of decaying solution must be
negligible \cite{8}.

  It is possible for the large perturbations that we see today to have
grown from the small perturbation present in decoupling time.
Typically we have$(\frac{\delta \rho}{\rho})_{dec} < (10^{-2} - 10^{-3})$ \cite{9}.

We compare the reach of the non-linear regime for open system cosmology with
the same quantity for standard model, using the quotient given by:
\begin{equation}
\Theta = \frac{\delta_{+c}}{\delta_{+F}}  \, ,
\end{equation}
where the coefficient $F$ refers to the Friedmann model and $c$ is relative to
the model with particle production.
In usual Friedmann universe  the reach of the non linear regime of the density
contrast is realized in a relative recent time, comparing with the age for the universe.  
 Consequently,  a cosmological model with better
characteristics to represent the universe must be satisfied if $\Theta > 1$.
Taking into account the growing mode (75) and  the condition (76)
they imply that $\beta < 0$, for $\nu =1 $.  Then we conclude that the respective cosmological
model to the source $\Psi = 3n\beta H$ does not give a better results than the usual FRW
model, since that for negative values for $\beta $, the second law of
thermodynamics is violated.

%SUBSECAO
%
%
\subsection{Source proportional to the $H^{\frac{2}{\nu }}$ }
The source
$
\Psi = \alpha H^{\frac{2}{\nu }}\, 
$
 is a natural generalization of the Prigogine's work \cite{14} for any value
of $ \nu $, excepted zero, where $\nu $ is a parameter of the state equation (67).

The field equations for the homogeneous and isotropic space time when particle
production is taking into account  can be written as:
\begin{eqnarray}
\rho = \frac{3}{k } H^2 \, , \\
\rho = \frac{\dot{n}}{n}(\rho + P_{th}) \, .
\end{eqnarray}
Using the state equation (67), the above field equations and the considered source we
find for the scale factor:
\begin{equation}
R(t) = \{1+a(e^{bt}-1)\}^ {\frac{2}{3\nu }}  \, , 
\end{equation}
where
\begin{eqnarray}
a = \frac{3}{\alpha }n_{0}^{\frac{\nu }{2}}(\frac{kM}{3})^{\frac{1}{3}} \, , \\
b = \frac{\alpha \nu }{2}(\frac{kM}{3})^{\frac{1}{\nu }} \, .
\end{eqnarray}
 The initial particle number density is denoted by $n_0 $ and $M$ is the mass
 of the particle
produced.

Substituting the scale factor in equation (64) and  changing the variable of the coordinate system using the following transformation
\begin{equation}
\xi = ae^{bt} \, ,
\end{equation}
we find:
\begin{eqnarray}
&\xi^2s(t)^{''}&+\xi (1+\frac{1}{3\nu })\{6+3\nu +\frac{4\xi }{\xi +1 -a}
\} s(t)^{'} + \{ \frac{2}{3(\xi +1 -a)} \\
&[3(a-1)& +\frac{4\xi }{\nu ^2}+\frac{3
  \xi}{\nu } -4\frac{a-1}{\nu }] +4\frac{(a-1)^2}{3\nu (\xi +1 -a)^2} \}s(t) =
0 \, . \nonumber
\end{eqnarray}
The derivatives are now taking in relation to the variable $\xi $.
The solution of last equation is given by the hypergeometric functions:
\begin{eqnarray}
&s(\xi )& = \{ C_{1}F_{1}(h1,h2,h3;\frac{\xi }{a-1}) +
C_{2}F_{2}(h4,h5,h6;\frac{\xi }{a-1})\\
&(\frac{\xi }{a-1})^{1-h4}&   \nonumber
 \} \xi ^{-\frac{3\nu ^2 +2 +6\nu -h1^{\frac{1}{2}}}{2\nu }}(\xi +a-1)^{-\frac{4+9\nu
    -h2^{\frac{1}{2}}}{6\nu }} \, .
\end{eqnarray}
where:
\begin{eqnarray}
&h_1& =\frac{1}{6\nu }\{ 3\,\nu+\sqrt {1080\,{\nu}^{2}+540\,{\nu}^{3}+528\,\nu+81
\,{\nu}^{4}+4} \\
&+&3\,\sqrt {9\,{\nu}^{4}+56\,{\nu}^{2}+36\,{\nu}^{3}+4+8 \nonumber
\,\nu}+\sqrt {16+24\,\nu+81\,{\nu}^{2}}\}
\end{eqnarray}
\begin{eqnarray}  
&h_2& = \frac{1}{6\nu }\{3\,\nu-\sqrt {1080\,{\nu}^{2}+540\,{\nu}^{3}+528\,\nu+81
\,{\nu}^{4}+4} \\
&+& 3 \sqrt {9\,{\nu}^{4}+56\,{\nu}^{2}+36\,{\nu}^{3}+4+8 \nonumber
\,\nu}+\sqrt {16+24\,\nu+81\,{\nu}^{2}}\}
\end{eqnarray}
\begin{equation}
h_3 = {\frac {\nu+\sqrt{9\,{\nu}^{4}+56\,{\nu}^{2}+36\,{\nu}^{3}+4+8\,\nu}}{\nu }} 
\end{equation}
\begin{eqnarray}
&h_4& = \frac{1}{6\nu }\{3\,\nu-\sqrt {1080\,{\nu}^{2}+540\,{\nu}^{3}+528\,\nu+81
\,{\nu}^{4}+4} \\
&-& 3\,\sqrt{9\,{\nu}^{4}+56\,{\nu}^{2}+36\,{\nu}^{3}+4+8  \nonumber
\,\nu}+\sqrt {16+24\,\nu+81\,{\nu}^{2}} \} 
\end{eqnarray}
\begin{eqnarray}
&h_5& =\frac{1}{6\nu }\{3\nu-\sqrt {1080\,{\nu}^{2}+540\,{\nu}^{3}+528\,\nu+81 \nonumber
\,{\nu}^{4}+4} \\
&-& 3 \,\sqrt{9\,{\nu}^{4}+56\,{\nu}^{2}+36\,{\nu}^{3}+4+8\,\nu} \\ \nonumber
&+&\sqrt {16+24\,\nu+81\,{\nu}^{2}} \}
\end{eqnarray}

\begin{equation}
h_6 = \nu-{\frac {\sqrt {9\,{\nu}^{4}+56\,{\nu}^{2}+36\,{\nu}^{3}+4+8\,\nu}}
{\nu}} \, .
\end{equation}

The coefficients of the  hypergeometrics usually obey the following relations:
\begin{eqnarray}
h_4= h_2 +1 -h_3 \, , \\
h5 = h_1 +1 -h_3 \, .
\end{eqnarray}
These conditions are valid for any value of $\nu $.
A third condition
\begin{equation}
h_3 = 2-h_6 \, .
\end{equation}
is valid only for $\nu =1$.
By inspecting the argument of the hypergeometric function, one is convinced
 very quickly that no easy analytical solution is possible.  In order to get
 an insight about the behavior of the evolution of the density contrast, we
 consider the constant  $a \approx 1 $,
consequently:
\begin{equation}
\xi \rightarrow \xi -1 +a \, .
\end{equation}
Then, the equation (83) assumes the form:
\begin{equation}
\xi ^2 s(\xi )^{''} +\xi (1+\frac{1}{3\nu })(10 + 3\nu )s(\xi )^{'} + 2\frac{4+3\nu }{3
  \nu ^2} s(\xi ) = 0 \, .
\end{equation}
Solution of the last equation is given by:
\begin{equation}
s(\xi ) = C_{1} \xi ^{w1-w2} + C_{2}\xi ^{w1+w2}
\end{equation}
where
\begin{equation}
w1 = -\frac{9\nu ^2 + 10 +30\nu}{6 \nu}
\end{equation}
and 
\begin{equation}
w2 = -\frac{\sqrt {81\nu ^4 +1080 \nu^2 +540\nu ^3 + 4 + 528\nu }}{6 \nu }
\end{equation}
When $t\longrightarrow 0 $, then $\xi \longrightarrow 1 $. So using the initial
condition $s(0) = 0 $,  the integration constants must obey the relation $C_{1}
= -C_{2} $. For mathematical convenience we take $ C_{1} = \frac{3}{2}$ .

Taking the solution for $s$, 
considering the parameter $ \tau = b t $ and $\nu = 1$,
the anisotropy of density contrast correspondent to the anisotropy of the cosmic background
 radiation of the microwaves ($s < 10^{-3} $) \cite{9} will  occur in the interval 
\begin{equation}
\tau _{d}  <  4.1 \, \times 10^{-5} \, ,
\end{equation}
for $\nu = 1 $.
When $\tau $ assumes the value
\begin{equation}
\tau _{l} \, \approx 7.3 \,\times  10^{-2} \, ,
\end{equation}
the density contrast reaches the non linear regime.
Now if we take $t$ instead of $\tau$ and considering that the anisotropy of CMBR
occurs at $\approx 10^{6} $ years, then the structure formation occurs at
\begin{equation}
t < 1.8 \, \times 10^9 \, ys \, .
\end{equation}

The density contrast in the  FRW standard model reach the non-linear regime at
$t< 3.16 \times 10^9 ys$, consequently in open system cosmology, for the
source considered, the structures
at large scale appeared in the universe more soon than in the standard model.
\section{Conclusions}
We obtain the evolution equation for the density contrast with respective solutions in the open system
cosmology (OSC).  

According to our results one may say that the OSC is an alternative model to
describe the universe.  The inclusion of the creation parameters indicates an
older universe \cite{Campos1} and furnishes a negative
pressure that accelerate the expansion  of the universe, according the recent
results obtained by the observations of the supernova.
We
analyze the conditions imposed on the creation parameters considering the
reach of the non-linear regime of the growing mode in a suitable time.
  The reach of the non-linear regime in the standard model  occurs in a more
  recent time than in the open system cosmology, considering the source of
  particle production given by $H^{\frac{2}{\nu}}$.
%\section{Observacoes}
%{\it gr-qc0002051-Acoustics of early universe}
%Classical perturbation theory formulated half a century ago by Lifshitz and
%Khalatnikov [] has nowadays been replaced by more appropriate gauge invariant
%descriptions[].
%
%Lifshitz theory provides the two parameter family of increasing  solutions for
%the density contrast(formula 115.19 of the Classical Theory of Fields), while
%all the gauge invariant approaches foresse in concert only a single growing
%density mode.
%


\begin{thebibliography}{99}

\bibitem{0}
V.F.Mukhanov,H.A.Feldman and R.H.Brandenberger,
{\it{Phys. Rep.}} {\bf 215 }, 203-333 (1992).

\bibitem{1}
J.M.Bardeen,
{\it{Phys. Rev.}} {\bf D22 },  1882 (1980).

\bibitem{2}
G.R.F, Ellis and M.Bruni,
{\it{Phys. Rev.}} {\bf D53 },  4287 (1989).

\bibitem{3}
J.A.Lima and A.S.Germano,
{\it phys.Lett} {\bf 170A}, 373,(1992)

\bibitem{4}
Winfried Zimdhal, Diego Pavon and David Jou,
{\it Class.Quantum Grav.} {\bf  8}, 677,(1993)

\bibitem{5}
V.N.Lukash,
{\it Sov.Phys-JEPT Lett} {\bf 12}, 307,(1980)
 
\bibitem{Kodama}
H.Kodama and M.Sazaki,
{\it Prog.Theor.Phys.Suppl },{\bf 78},1, 1984 

\bibitem{14}
I.Prigogine, J.Geheniau, E.Gunzig and P. Nardone,
{\it GRG} {\bf 21}, 8, 767.

\bibitem{13}
M.O.Calv\~ao,J.A.S.Lima and I.Waga,
{\it phys.Lett} {\bf 162A}, 223,(1992).

\bibitem{6}
J.A.S.Lima, A.S.M.Germano and L.R.W.Abramo,
{\it{Phys. Rev.}} {\bf D53 },  4287 (1996).

\bibitem{12}
R.M.Wald, {\it General Relativity} (University of Chicago Press, Chicago).

\bibitem{7}
S.Weinberg, {\it Gravitation and Cosmology: Principles and Applications of  the General Theory of  Relativity}  (Jonh Wiley \& Sons, New York, 1972). 

\bibitem{8}
J.Peebles et al.,
{\it ApJ} {\bf 147}, 859,(1967).

\bibitem{9}
E.W.Kolb and M.S. Turner,
{\it The Early Universe } (Addison-Wesley ,Redwood City,California,1990) . 


\bibitem{Campos1}
 M. de Campos and N. Tomimura,
{\it astro-ph/9911276 } {\bf 494 }, 96 (1998)

\end{thebibliography}
 \end{document}